\documentclass[aps,prl,preprint,superscriptaddress,showpacs,showkeys]{revtex4}

\usepackage{graphics}
\usepackage{longtable}
\usepackage{natbib}
\begin{document}

\title{On folding and form: Insights from lattice simulations}

\author{P.F.N. Faisca}
\affiliation{CFTC, Av. Prof. Gama Pinto 2, 1649-003 Lisboa Codex, Portugal}
\email{patnev@alf1.cii.fc.ul.pt}
\author{M.M. Telo da Gama}
\affiliation{CFTC, Av. Prof. Gama Pinto 2, 1649-003 Lisboa Codex, Portugal}
\author{R.C. Ball}
\affiliation{Department of Physics, University of Warwick, Coventry
CV4 7AL, U.K.}

\date{\today}

\begin{abstract}
Monte Carlo simulations of a Miyazawa-Jernigan lattice-polymer model indicate that, depending on the native structure's geometry, the model exhibits two broad classes of folding mechanisms for two-state folders. Folding to native structures of low contact order is driven by backbone distance and is characterised by a progressive accumulation of structure towards the native fold.
By contrast, folding to high contact order targets is dominated 
by intermediate stage contacts not present in the native fold, yielding a 
more cooperative folding process.  
\end{abstract}
\pacs{87.14.Ee; 87.15.Aa}
\keywords{protein folding, contact order, kinetics}
\maketitle

\section{Introduction}
Advances in experimental techniques and the use of
computational models have shown that most small 
(from $\sim$ 50-120 amino acids), 
single domain proteins fold via two-state kinetics, 
without observable folding intermediates and with a single transition state 
associated with one major free energy barrier separating the native from the 
unfolded conformations~\cite{JACKSON,PLAXCO0,FER,PFN,KAYA1}.
In addition it is also well known that two-state
proteins, with similar chain lengths, exhibit a remarkably wide range
of folding rates, folding in microseconds to seconds 
~\cite{PERNILLA,NIKOLAY,PLAXCO}. 
Understanding what makes some proteins such incredibly fast
folders will shed light into the underlying folding mechanism. \par
The energy landscape theory predicts that the landscape's ruggedeness
plays a fundamental role in the folding kinetics of proteins: 
The existence of local energy minima, that act as kinetic traps, is 
responsible for the 
overall slow and, under some conditions (as the temperature
approaches the glass transition temperature), glassy dynamics.
However, and as pointed out by Du {\it {et al.}}~\cite{DU}, another equally
important aspect of the folding dynamics is related to the 
geometry of protein chain conformations. Indeed, both chain connectivity and
(steric) excluded volume impose restrictions on the number of allowable 
conformations a polypeptide chain can adopt and these geometrical constraints 
play a significant role in determining the folding pathways that are 
kinetically 
accessible.\par
A quantitative measure of geometric complexity, the so-called relative contact
order, $CO$, was introduced in 1998 by Plaxco {\it{et al.}}~\cite{PLAXCO}:
The $CO$ is a simple, empirical parameter measuring
the average sequence separation of contacting residue pairs
in the native structure relative to the chain length of the protein
\begin{equation}
CO=\frac{1}{LN}\sum_{i,j}^N \Delta_{i,j}\vert i-j \vert,
\label{eq:no1}
\end{equation}  
where $\Delta_{i,j}=1$ if residues $i$ and $j$ are in contact and is 
0 otherwise; $N$ is the total number of contacts and $L$ is the protein chain 
length.
A strong correlation ($r=0.94$) was found between the $CO$ and the 
experimentally observed folding 
rates in a set of 24 non-homologous single domain proteins~\cite{PLAXCO2} 
suggesting a topology-dependent kinetics of two-state folders. 
Results obtained by two of us~\cite{PFN2} in the context of a simple 
Miyazawa-Jernigan (MJ) lattice polymer model~\cite{PFN2}
showed a significant correlation ($r=0.70-0.79$) between increasing $CO$ and the 
longer logarithmic folding times. In a more recent
study Jewett {\it {et al.}}~\cite{JEWETT} found a similar corelation
($r=0.75$) for a 27-mer lattice polymer modeled by a modified G\={o}-type 
potential. These results support the empirical correlation found
between contact order and the kinetics of two-state folders.\par
In this paper we investigate wether the geometry of the native structure 
does, or does not, promote different folding processes, eventually leading 
to different folding times, in the context of the MJ 
lattice-polymer model and Monte Carlo (MC) folding simulations. 
Although lattice models are not capable of describing the full complexity 
of real proteins they are non-trivial and thus may capture some 
fundamental aspects of protein folding kinetics~\cite{SFERMI}.  
The native structures considered in our study were selected on the basis 
of their different 
contact order parameters. The $CO$ is clearly not the only way to quantify 
the native 
structure's geometry but the empirical finding that the $CO$ correlates well 
with the folding rates of real proteins strongly motivates its use for the 
purposes of the present work.\par
The paper is organized as follows: Section II reviews the model and methods 
used in the lattice simulations. In section III the numerical results are 
presented. We start with a preliminary study emphasizing the gross 
distinctive features observed between the folding dynamics to low and to 
high-$CO$ structures. Subsequently we make a more detailed 
analysis of the folding dynamics associated with the low-$CO$ and high-$CO$ 
native structures that exhibit respectively the lowest and the highest 
folding times in order to highlight specific traits of the 
respective folding processes. In section IV we make some final remarks and 
summarize our conclusions.          

\section{Model and methods \label{sec:secno2}}
We consider a simple three-dimensional 
lattice model, based on a bead and stick representation, of a protein molecule.
In such a coarse grained model a bead represents an amino acid and the unit 
length stick stands for the peptide bond that covalently connects amino acids 
along the polypeptide chain.
The chains consist of $N=48$ beads interacting via short-range interactions 
described by the contact Hamiltonian
\begin{equation}
H(\lbrace \sigma_{i} \rbrace,\lbrace \vec{r_{i}} \rbrace)=\sum_{i>j}^N
\epsilon(\sigma_{i},\sigma_{j})\Delta(\vec{r_{i}}-\vec{r_{j}}),
\label{eq:no2}
\end{equation}
\noindent
where $\lbrace \sigma_{i} \rbrace$ represents an amino acid sequence,
$\sigma_{i}$ standing for the chemical identity of bead $i$, while 
$\lbrace \vec{r_{i}} \rbrace$ is the set of bead coordinates defining 
each conformer.
The contact function $\Delta$ is $1$ if beads $i$ and $j$ are in
contact but not covalently linked and is $0$ otherwise. We follow previous 
studies~\cite{SG, SG2, SALI, TIANA, JMBS} by taking the interaction parameters 
$\epsilon$ from the $20 \times 20$
MJ matrix, derived from the distribution of contacts in native
proteins~\cite{MJ}. \par
The folding simulations follow the standard MC Metropolis 
algorithm~\cite {METROPOLIS} and, in order to mimic protein movement, 
we use the kink-jump MC move set, including
corner flips, end and null moves as well as crankshafts~\cite{BINDER}.\par
Each MC run starts from a randomly generated unfolded conformation 
(typically with less than
10 native contacts) and the folding dynamics is traced by following the 
evolution of the 
fraction of native
contacts, $Q=q/Q_{max}$, where $Q_{max}=57$ and $q$ is the number of native 
contacts at each 
MC step.
The folding time, $t$, is taken as the first passage time (FPT), that is,
the number of MC steps that corresponds to $Q=1.0$.

\section{Numerical results \label{sec:secno3}}
\subsection{Targets}
The distribution of the relative contact order parameter over a population of 
500 
target geometries, folding to fill a simple cuboid, 
was found via homopolymer relaxation~\cite{PFN2} and exhibits $CO$ values 
that span the 
intervals centered around $CO=0.13$ and $CO=0.26$.
To investigate the effects of $CO$ on the folding dynamics we selected
from our target pool the three lowest-$CO$ and the three highest-$CO$ 
maximally compact structures as the targets of our protein model.\par

\subsection{A preliminary analysis of the folding dynamics}
For each target, a set of 100 proteinlike sequences was prepared using the  
Shakhnovich and Gutin design method~\cite{SG}. 
The averaged trained sequence energy, $<E>$, and its standard
deviation, $\sigma$, are shown in Table~\ref{tab:tabno1}, where the targets 
are ordered with increasing $CO$.\par
The folding dynamics was studied at the so-called optimal folding temperature, 
the temperature that minimizes the folding time, taken as the value of the 
mean FPT to the target
averaged over the 100 MC runs~\cite{PFN2}. Note that the high-$CO$ targets 
are associated with folding times that are systematically larger than those 
associated with the low-$CO$ targets. 
Indeed, in this 6-element target set, contact order and folding times
correlate well ($r=0.84$).
The simulated range of folding rates is, however, much narrower
than that observed in real proteins ($\approx$ 5 orders of magnitude); 
the simulated  kinetics is typical of this type of models and thus it 
appears to be a limitation of the lattice polymer model as well as of some 
continuum, off lattice, models that exhibit similar behaviour~\cite{KAYA1}.\par
In order to trace conformational changes we used the so-called contact 
map ~\cite{CMAP}.
The contact map, ${\bf C}$, is an $N \times N$ symmetric matrix with elements, 
$C_{ij}=1$, if beads $i$ and $j$ are in contact (but not covalentely 
linked) and zero otherwise.
In addition to containing the relevant information on the protein's 
structure (total number of contacts, specification of each contact and 
respective range) the contact map representation provides a straightforward 
way to compute the frequency, $\omega_{ij} =t_{ij}/t$, with which a native 
contact $ij$ occurs in a MC run, 
$t_{ij}$ being the total number of MC steps where
$C_{ij}=1$ and $t$ the folding time.
We have grouped the contacts into two classes, based on their frequency: If 
$\omega \geq 0.5$ the contact is long-lived, while short-lived contacts are 
those with a frequency $0.4 \le \omega < 0.5$. We have focussed on
the contacts that contribute to the folding process and thus excluded 
from the analysis contacts with small or marginal lifetimes.\par
We computed the mean frequency of each native contact $<\omega_{ij}>$, 
averaged over 100 simulation runs, and report the results on 
Table~\ref{tab:tabno1}.
We note that in the low-$CO$ set, the fraction of native contacts with a 
significant 
lifetime, is approximately twice as large as the corresponding fraction in 
the high-$CO$ 
set. In both sets, however, most of the long-lived contacts are local 
(a contact is local if the contacting beads are separated by less than 10 
units of backbone distance), possibly due to the local nature of the 
kink-jump dynamics move set. \par
By contrast, the fraction of short-lived contacts is similar in both target 
sets; naturally the number of long-range (LR) contacts, contributing to $Q$, is 
clearly larger in the high-$CO$ target set. 
The number of non-native contacts, $N_{nnat}$, with a marginal lifetime 
($\omega \geq 0.10$) is, as expected, larger in the high-$CO$ target set.\par
These results indicate that the fraction of long-lived native 
contacts is higher in chains folding to low-$CO$ targets and that, 
regardless 
of target geometry, the dynamics appears to be dominated by local contacts 
as these are the most frequent. Nevertheless, the 
appearance of a few long-lived LR contacts in both target sets suggests that 
they may play a role in the folding dynamics of these proteins.
\par 

\subsection{Contact order and structural organization towards the native fold}
In this section a detailed study of the folding dynamics exhibited by
targets $T_{1}$ and $T_{5}$ is investigated. 
Targets $T_{1}$ and $T_{5}$ have considerably different
geometries, as suggested by their contact order, and display the 
lowest and the highest observed folding times. Therefore they are good 
candidates to highlight the role of the native structure's geometry (if 
any) on the folding dynamics.  
In particular, we investigate wether specific structural changes 
towards the 
native fold may be identified, for a given native structure's 
geometry.\par
\begin{figure}
\centerline{\rotatebox{270}{\resizebox{8.6cm}{8.6cm}{\includegraphics{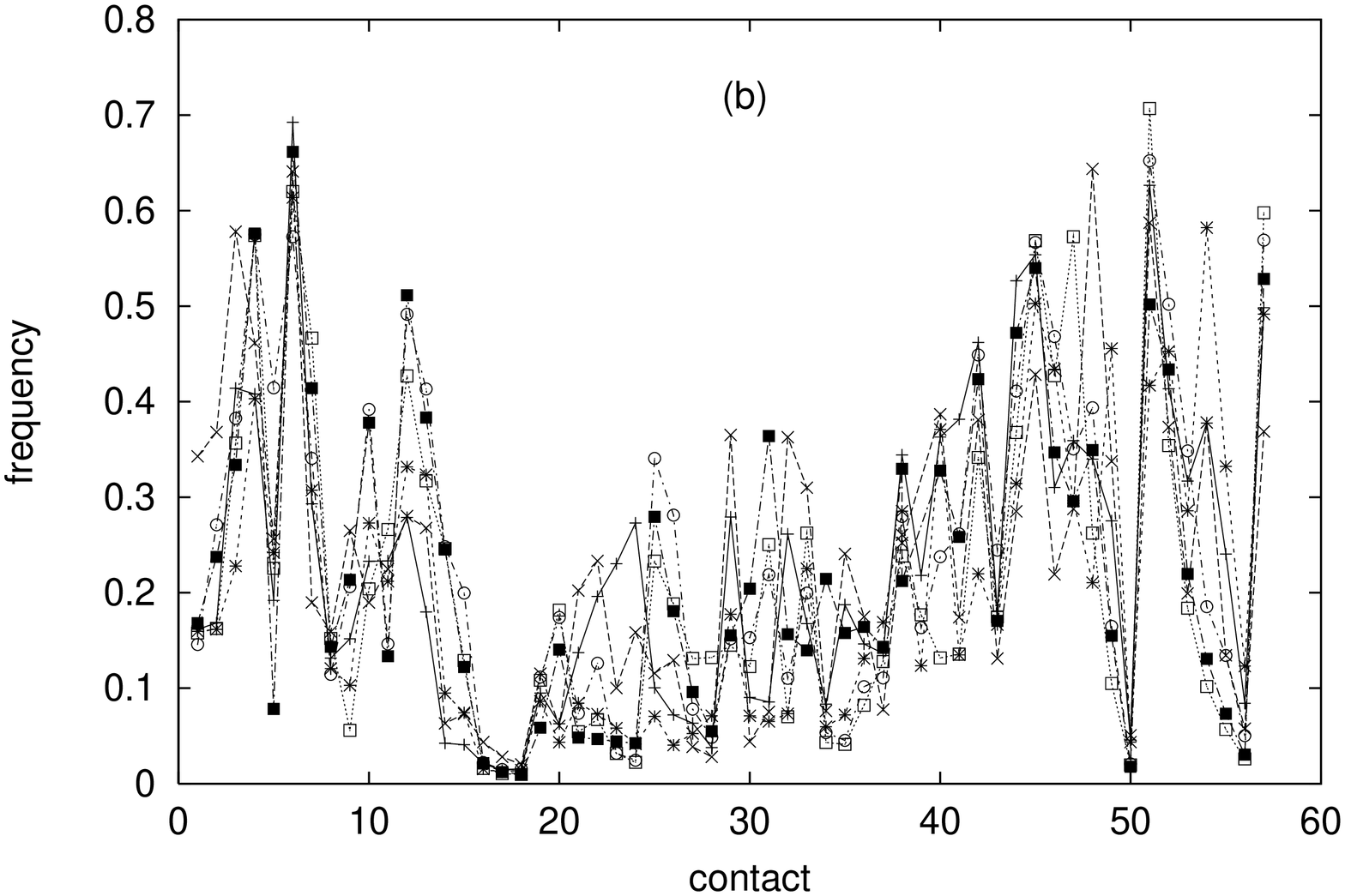}}}}
\caption{Frequency with which a native contact 
(numbered from 1 to 57) occurs in the
folding simulations of six randomly chosen sequences trained for targets
$T_1$ (a) and $T_5$ (b). The contact frequency is the ratio of the number of times a native contact occurs in a MC simulation to the folding time. Note how frequency of occurence of particular contacts has strong correlation between different trained sequences; that is a clear dependence on conformation alone.
}
\label{figure:no1}
\end{figure}

In Fig.~\ref{figure:no1} we plot the frequency, $\omega_{ij}$, with which a 
native contact $ij$ appears in the folding simulations of six randomly chosen
sequences trained for targets $T_{1}$ and $T_{5}$ respectively. 
The major features observed for each target in different runs suggest 
a trend for the folding dynamics of target $T_{1}$ that is markedly different 
from that observed for target $T_{5}$. In what follows we will investigate this 
difference.\par
\begin{figure}
\rotatebox{270}{\includegraphics{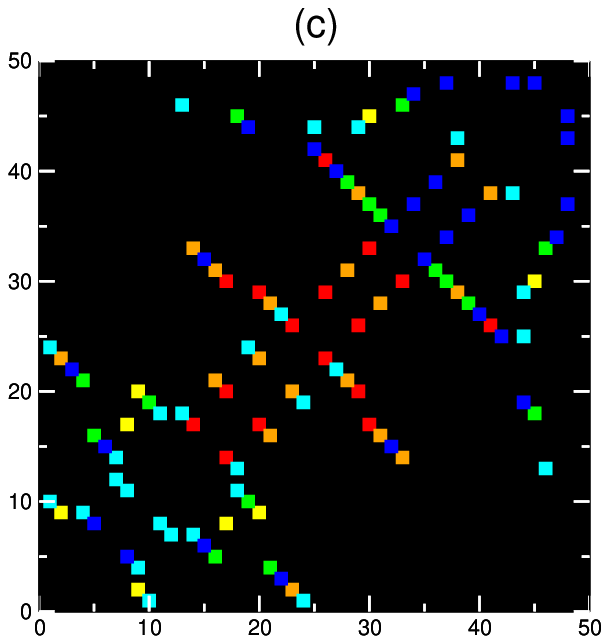}}
\rotatebox{270}{\includegraphics{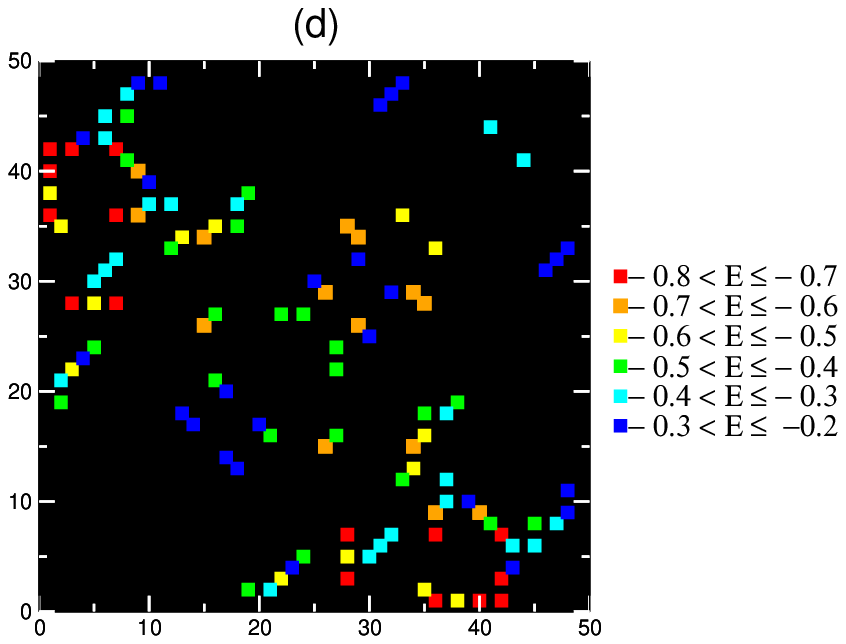}}
\caption{(Color online). Frequency maps of targets $T_1$ (a) and $T_5$ (b) and
energy maps of targets $T_1$ (c) and $T_5$ (d). A coloured square represents a native contact with an averaged mean frequency $<\omega>$ or
an averaged mean energy $E$. Averages are taken over 100 MC runs.}
\label{figure:no2}
\end{figure}

Figures~\ref{figure:no2}(a) and (b) show the {\it frequency maps} of targets 
$T_{1}$ and $T_{5}$ respectively. Each square represents an element 
$C_{ij}=1$ of the contact map matrix, that is, a native contact $ij$, whose 
mean frequency, $<w_{ij}>$, averaged over 100 MC runs, falls in a certain 
range indicated by the different colours. The frequency maps 
clearly identify the two model structures $T_{1}$ and $T_{5}$ and exhibit 
their different geometries. It is possible to identify a pattern in the colour 
distribution of target $T_{1}$, which is not present in the frequency map of 
target $T_{5}$, suggesting that the mean frequency of a native contact 
decreases monotonically with increasing contact distance in the low-$CO$ 
target. \par
\begin{figure}
\centerline{\rotatebox{270}{\resizebox{8.6cm}{8.6cm}{\includegraphics{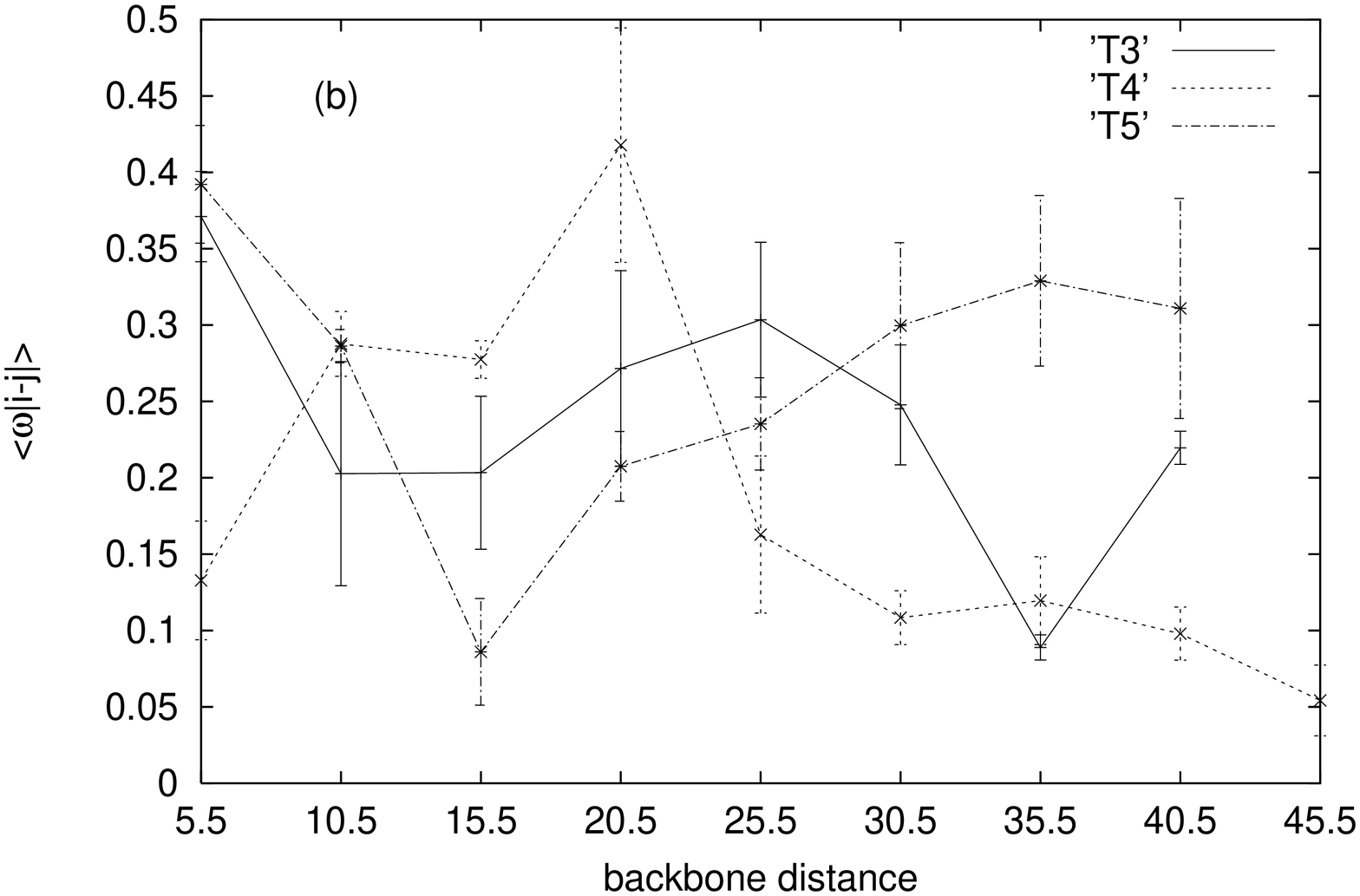}}}}
\centerline{\rotatebox{270}{\resizebox{8.6cm}{8.6cm}{\includegraphics{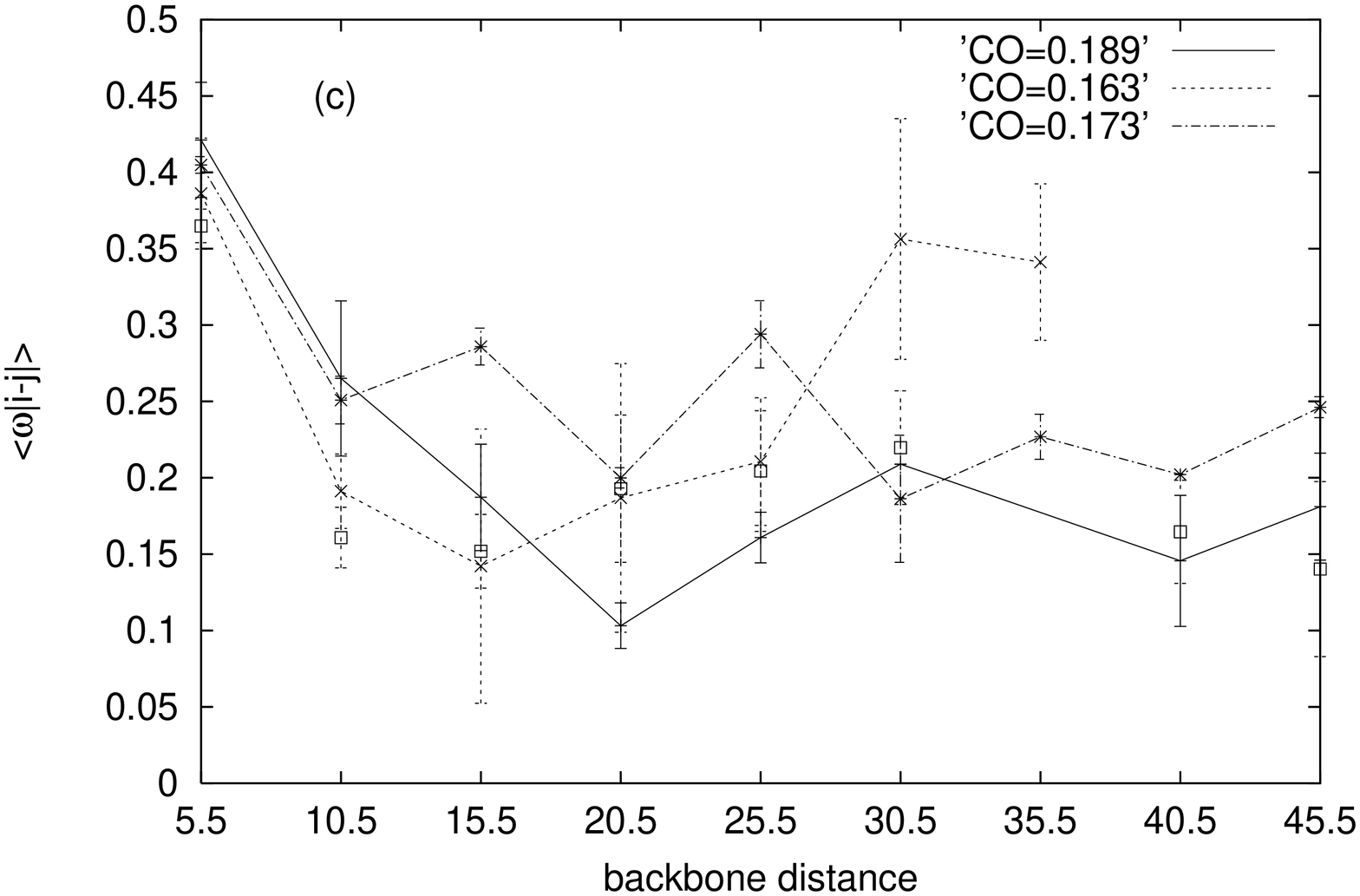}}}}
\caption{The backbone frequency, $<\omega_{\vert i-j \vert}>$ , as a function of the backbone separation 
for the low-$CO$ (a), high-$CO$ (b) and intermediate-$CO$ targets (c). The backbone frequency is the mean value of $<\omega>$ averaged over the number of contacts in each interval of backbone separation as shown in Table~\ref{tab:tabno2}.} 
\label{figure:no3}
\end{figure}

Let the backbone frequency, $<\omega_{|i-j|}>$, be the mean frequency 
$<\omega_{ij}>$
averaged over the number of contacts in each interval of backbone separation 
as defined in Table~\ref{tab:tabno2}. 
In Figs.~\ref{figure:no3}(a) and ~\ref{figure:no3}(b) we plot the backbone 
frequency as a function of the distance, $|i-j|$ for the targets of the 
low-$CO$ and the high-$CO$ sets respectively. While for all 
low-$CO$ targets $<{\omega}_{|i-j|}>$ decreases monotonically with 
increasing contact distance, confirming
the trend observed in the $T_{1}$ frequency map, for the high-$CO$ targets 
no such trend is observed. One possible explanation, 
that we have ruled out, is that of a (negative) correlation
between the frequency of a contact and its energy. In particular, one 
might expect the most stable contacts, those with the lowest 
energy, to be the most frequent. 
In Fig.~\ref{figure:no2}(c) and (d) we report the {\it energy maps} of 
targets $T_{1}$ and $T_{5}$ respectively. Each square represents
a native contact whose mean energy, averaged over 100 sequences,
falls in a range indicated by the colour.
Since there is no correspondence between the colour patterns of
figures~\ref{figure:no2}(a) and ~\ref{figure:no2}(c) and between those of 
figures~\ref{figure:no2}(b) and ~\ref{figure:no2}(d)  
we conclude that the difference is driven by 
geometrical constraints. A quantitative analysis of the correlation between 
the contact's frequencies and energies yields modest correlation 
coefficients, $r=0.63$ and $r=0.65$ for targets $T_{1}$ and $T_{5}$ 
respectively.\par
\begin{figure}
\centerline{\rotatebox{270}{\resizebox{8.6cm}{8.6cm}
{\includegraphics{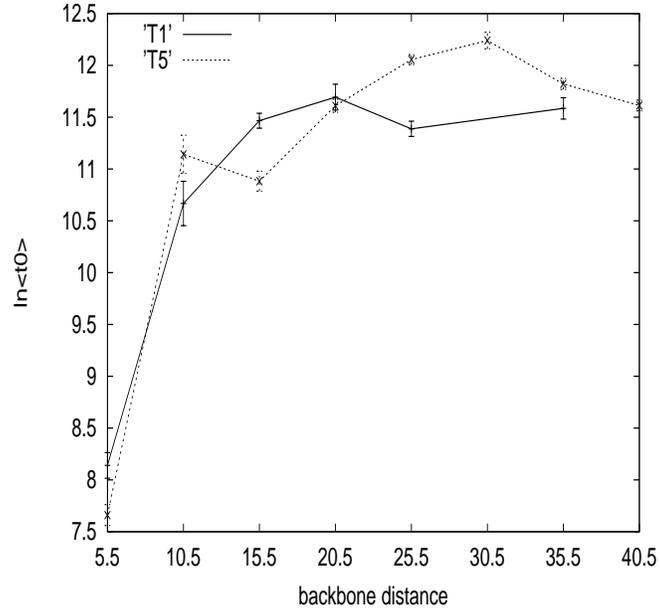}}}}
\caption{The averaged contact time, $\ln_{e}<t_{0}>$, as a function of the backbone distance. $t_{0}$ was averaged over the number of contacts in each interval of backbone separation as shown in  Table~\ref{tab:tabno2}.}
\label{figure:no4}
\end{figure}

Let the contact time, $t_{0}$, be the mean FPT of a native contact 
averaged over 100 MC runs (the FPT of a native contact is number of MC 
steps up to the first time the contact is formed).
The contact time, averaged over the contacts in each interval of 
backbone distance, are shown in Table~\ref{tab:tabno2}, and plotted 
in Fig.~\ref{figure:no4}: In both targets the 
set up of local contacts occurs largely before the LR
contacts are established and, for LR contacts, there is no correlation
between the contact time and the backbone distance.
In view of these results one may be tempted to conclude that the higher 
folding time of $T_{5}$ is due to the fact that it has more LR 
contacts. However, the folding time is non-additive and a simple 
calculation shows that the higher number of LR contacts cannot justify the 
observed folding time of $T_{5}$. Indeed the longest contact time 
($\ln t_0=12.24$) is two orders of magnitude shorter than the  folding 
time of  $T_{5}$ and the sum of contact times is $\ln (\sum_{i=1}^{57} t_{0}^{i})=15.51$ much lower than the observed folding time $\ln t=17.59$. 
\par
From the results of Fig.~\ref{figure:no4} we infer that the average 
contact times, over a given range, are similar for both targets. Thus the 
differences 
in the observed frequencies reported in Fig.~\ref{figure:no3} distinguish 
different cooperative behaviours. \par
Results obtained so far suggest that two broad classes of folding mechanisms
exist for the MJ lattice-polymer model. What 
distinguishes these two classes is the presence, or absence, of a 
monotonic decrease of contact frequency with increasing contact range
that may be related to different cooperative behaviour.
The monotonic decrease of contact frequency with increasing contact range
appears to be specific of the folding to low-$CO$ targets. 
In this case the folding is also less cooperative and seems to be driven 
by the backbone 
distance: Local contacts form first while LR contacts form progressively 
later as the contact range increases. \par
At this point one may ask if the different folding mechanisms identified 
in the previous discussion are not a consequence of analysing only 
two different structures, i.e. there could be intermediate mechanisms for 
intermediate native structures. 
In order to clarify this point, we have analysed the folding of 
Shakhnovich and Gutin sequences 
designed to three target geometries with intermediate contact order 
(0.163, 0.173 and 0.189) and the results are reported in 
Fig.~\ref{figure:no3}(c). The folding times associated with these 
3 structures are $15.67\pm 0.09$, $16.46 \pm  0.09$ and $16.04 \pm 0.12$ 
and we found that the contact order and the folding times, for the 
9-element target set, correlate well ($r=0.82$). 
The average sequence energy is in the same range as that of the targets 
reported in Table~\ref{tab:tabno1}. However, it is clear from the figure 
that intermediate and high-$CO$ proteins fold via the same type of 
cooperative mechanism.\par

\subsection{Contact order and the exploration of the conformational space}

In this section we analyse the time evolution of the 57 native contacts of 
targets $T_{1}$ and $T_{5}$ to obtain a picture of the `global' structural 
changes that occur during folding. \par
In the folding process a chain explores conformations that may be 
characterised by the fraction of native contacts, $Q$. Different native 
contacts will contribute to conformations with the same $Q$. In a MC run 
the probability of occurrence of a certain native contact is equal to the 
number of times that the contact occurs over the number of times that 
conformations with a given fraction of native contacts, $Q$, are sampled.

\begin{figure}
\centerline{\rotatebox{0}{\resizebox{8.6cm}{8.6cm}{\includegraphics
{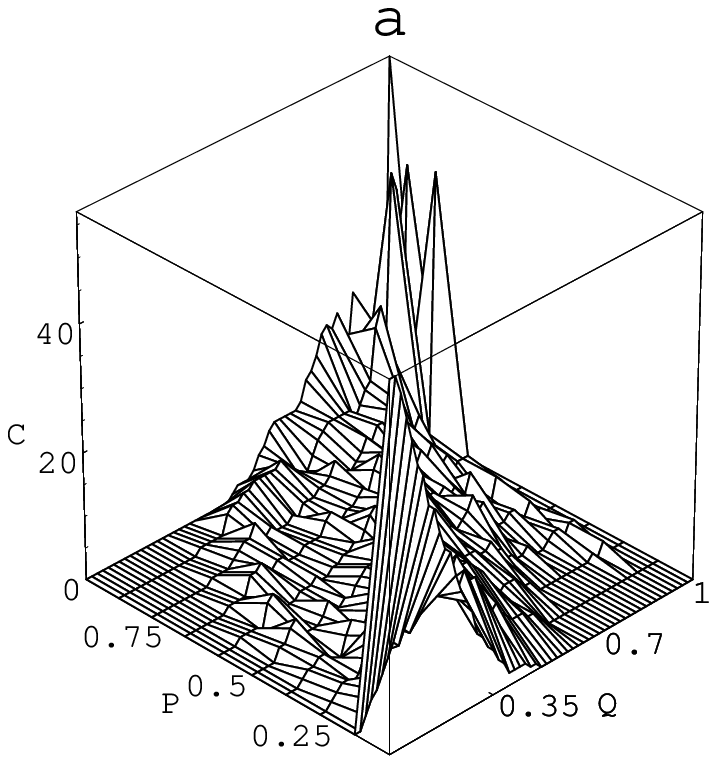}}}}
\centerline{\rotatebox{0}{\resizebox{8.6cm}{8.6cm}{\includegraphics
{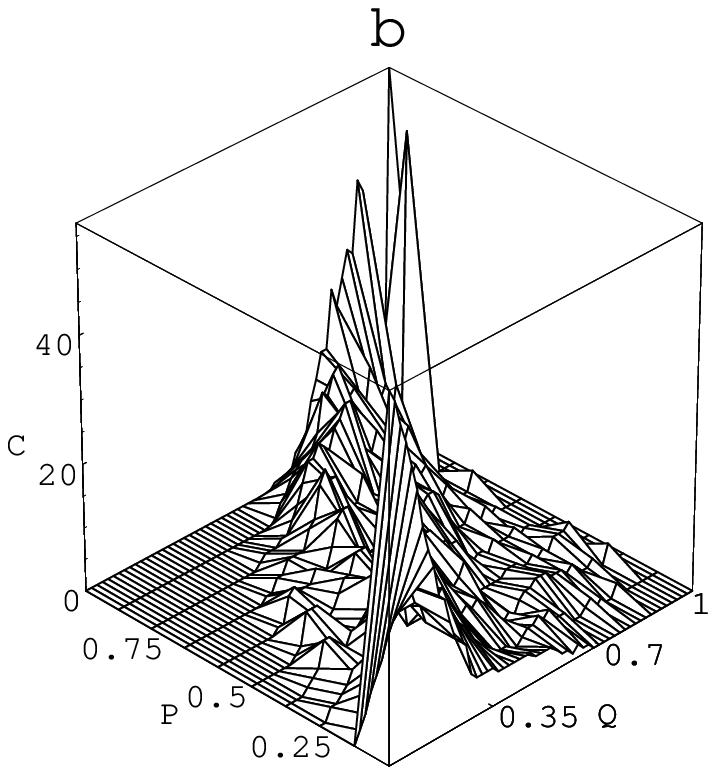}}}}
\caption{Number of contacts, $C$, with a given probability of being formed, $P$, as a function of $Q$, the fraction of native contacts for targets $T_{1}$ (a) and $T_{5}$ (b). These are results averaged over 100 MC folding runs.} 
\label{figure:no5}
\end{figure}

Since in a given run some native contacts are more probable (or more 
frequent) than others one may consider different probability intervals and 
ask, from the total number of native contacts, how many occur within a 
given probability interval, at fixed $Q$. 
The result gives the dependence of the number of contacts, $C$,
on $P$, the probability of a contact being formed, and on $Q$, the 
fraction of native contacts~\cite{KPLUS}.
Results, averaged over 100 simulation runs, are reported 
in Fig.~\ref{figure:no5}(a) for $T_{1}$ and in Fig.~\ref{figure:no5}(b) 
for $T_{5}$ where the coordinate $Q$ may be interpreted as a 
monotonic `time' coordinate along the folding process. Accordingly, early 
folding corresponds to low-$Q$ while late folding occurs at high-$Q$.\par
A first look at the figures suggests smooth dynamics for the time evolution 
of $T_1$'s 57 native contacts by comparison with $T_5$ that exhibits a 
considerably more `rugged' behaviour. 
Indeed, for a fixed probability interval, the variation, as a function 
$Q$, 
of the number of native contacts $C$ which are present with that probability 
is clearly more pronounced for $T_{5}$ than for $T_{1}$. This suggests that
$T_{5}$ does not keep a considerable number of its native contacts as it 
evolves from a conformation $\Gamma (Q)$ to another conformation 
$\Gamma^{'} (Q^{'})$ during its exploration of the conformational space 
towards the native fold. 
A closer look shows other important differences. In
the early folding ($Q<0.35$) of $T_1$ there are a few near 
permanent contacts, that is highly probable contacts ($P\ge 0.80$), by 
contrast with 
$T_5$ where highly probable contacts occur only later ($Q\ge 0.50$). 
Indeed, in the 
late folding of $T_{5}$, there are still a few contacts with rather low 
probability $P\approx$0.25.
Moreover, as $Q$ increases, the number of contacts in the two highest 
probability intervals increases smoothly for $T_{1}$, while for $T_{5}$ 
the number of high-probability contacts shows a sudden increase only at $Q 
\approx 0.7$.
These dynamical features are consistent with a folding scenario according 
to which $T_{1}$ explores more correlated native-like conformations as time 
evolves. For $T_{5}$, however, even though the chain is getting more compact
as it evolves towards the native fold it still explores 
many uncorrelated conformations up to the late folding stage. \par

\subsection{Contact order and non-native contacts}
\begin{figure}
\centerline{\rotatebox{270}{\resizebox{8.6cm}{8.6cm}{\includegraphics
{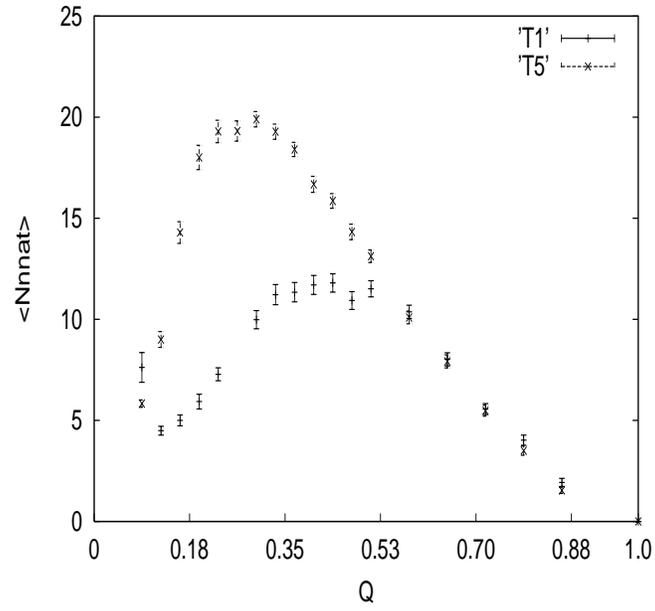}}}}
\caption{Mean number of non-native contacts, $<N_{nnat}>$, averaged over 100 MC runs, as a function of the fraction of native contacts.} 
\label{figure:no6}
\end{figure}

To investigate the effects of non-native contacts in the folding dynamics
to geometrically different native structures
we have computed the dependence of the averaged number of 
non-native contacts, 
$<N_{nnat}>$, with $Q$. The average is taken over 100 MC runs. Results 
reported in 
Fig.~\ref{figure:no6} show that it is possible to identify two distinct 
dynamical regimes: For $Q>0.5$ 
the number of non-native contacts decreases monotonically with $Q$ 
independent of target geometry.
However, for lower $Q$, the dynamics is target sensitive with 
the high-$CO$ target displaying a larger number of total contacts.
This data is consistent with a folding scenario where, in the early folding 
of the high-$CO$ target, 
conformational sampling is geometrically restricted due to 
pre-existing compact structures.\par
Within the context of the energy landscape theory significant energy barriers, 
or kinetic traps, are known to exist between compact denaturated structures 
slowing down the folding process. 
Can the observed compact structures act as kinetic traps in the folding 
of $T_{5}$? \par
\begin{figure}
\centerline{\rotatebox{270}{\resizebox{8.6cm}{8.6cm}{\includegraphics
{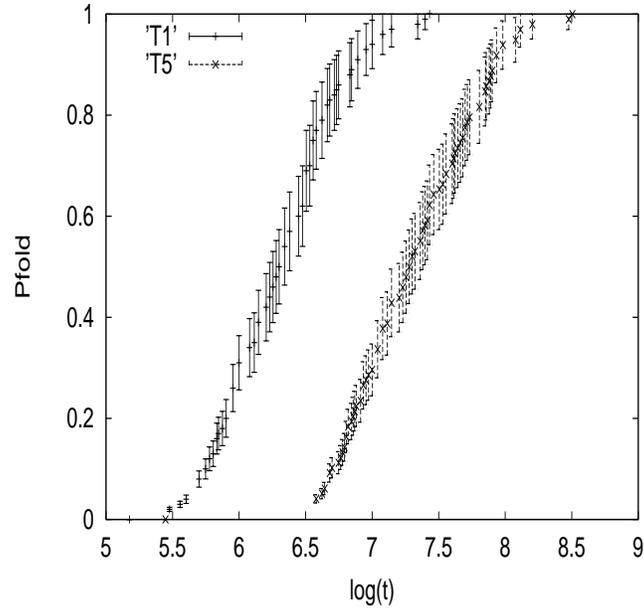}}}}
\caption{Dependence of the folding probability, $P_{fold}$, on $log_{10}(t)$.
$P_{fold}$ was calculated as the number of folding simulations which ended up to time $t$ normalized to the total number of runs.} 
\label{figure:no7}
\end{figure}

In order to answer this question we have
computed the transition probability curves for both targets where the
presence of plateaus indicates the presence of kinetic traps. In a 
transition probability curve
the folding probability, $P_{fold}(t)$, is plotted against $t$. 
Results for targets $T_{1}$ and $T_{5}$ are shown in 
Fig.~\ref{figure:no7} where no plateaus are visible. Thus,
based on these results, one cannot claim that the longer folding 
time of $T_5$ is due to the presence of kinetic traps.\par
Why do high-$CO$ structures form compact denatured states?
We associate these conformers with the existence of high-frequency, LR 
native contacts. Indeed, native contacts with a backbone separation in the 
range $35 \leq |i-j| \leq 39$ and  
frequency in the range $0.40 \leq \omega_{ij} \leq 0.57$ correspond to 
conformers characterized
by $Q \approx 0.18$ and a total number of contacts close to $30$ 
(of which $\approx 20$ are non-native). Figure~\ref{figure:no6} confirms 
that for $Q=0.18$ target $T_{5}$ is considerably more compact than target 
$T_{1}$ that has only  
($\approx 12$) non-native contacts. \par
These findings suggest the following interpretation of the
behaviour observed in Fig.~\ref{figure:no5} for the dynamics of the 
high-$CO$ target's ensemble of native contacts: The promiscuous formation of LR 
contacts takes the chain through low conformational entropy states
from where it reorganises in a time consuming process towards the native fold. 
This major 
reorganization explains why even in the late stages of folding the chain is 
still exploring sets of unrelated conformations.\par

\section{Conclusions and final remarks \label{sec:secno4}}

In the present work we have carried out a thorough statistical analysis of 
the folding dynamics of 48 mers, within the MJ lattice-polymer model, 
designed to high, intermediate and low-$CO$ target structures, in order to 
investigate the folding mechanisms associated with different 
target geometries, and the corresponding folding rates. 
\par 
We found two broad classes of folding mechanisms for the MJ 
lattice-polyme r model. The main feature of the 
first class, that describes the folding of low-$CO$ targets, is a monotonic 
decrease of contact frequency with increasing contact range; indeed, such 
dependence seems to be a specific trait of the dynamics associated with 
low-$CO$ targets. The building up of native structure is
driven by backbone distance with local contacts forming first and 
non-local 
contacts forming 
progressively later as contact range increases. Moreover, the analysis of the 
time evolution of the 57 native contacts shows a progressive cumulative 
construction of the native fold with the chain exploring more correlated 
native-like conformations as time evolves. 
Folding to low-$CO$ native structures is therefore gradual rather than 
abrupt (or 
cooperative). Folding to intermediate and high-$CO$ targets belongs to a 
different class, where the dependence of contact frequency on contact range 
is non-monotonic. The folding is markedly more cooperative
with many high-probability contacts forming suddenly only in the late 
stages of folding.
Our results suggest that the higher cooperativity of the high-$CO$ folding 
dynamics is due to the presence of LR contacts. A similar conclusion on 
the role of LR contacts in the folding dynamics was obtained 
by Abkevich {\it et al.} in Ref. {\cite{JMBS}}. \par 
A common feature of the two folding classes is that
the dynamics is dominated by local contacts in the sense that they are the 
most frequent during the folding process. This feature results, in part, 
from the local nature of the move set used in the simulations 
which favours the formation of local contacts. \par
At this stage a word on the correlation between $CO$ and foldind times is in 
order. Although the correlation coefficient between $CO$ and $T_f$ for the 
6 targets of Table I is high ($r=0.82$) the difference in folding times is 
relatively modest and this correlation should be taken with caution. 
Indeed, one one includes the 9 targets studied in this work the 
correlation coefficient decreases, a clear indication that these numbers 
are not conclusive. However, the geometry driven cooperativity appears to 
be rather robust and this implies an increase in folding times as the 
cooperativity increases. \par
Related studies, have investigated the physical mechanisms behind the 
(empirical) geometry-dependent kinetics exhibited by two-state folders.
Work on the `topomer search model' (TSM), concludes that the 
topology-dependence of real two-state folders is `a 
direct consequence of the extraordinary cooperative equilibrium folding
of simple proteins' {\cite{PLAXCO3}}. In agreement with the TSM results 
Jewett {\it et al.} {\cite{JEWETT}} showed that  
modified G\={o} type polymers, exhibiting enhanced thermodynamic cooperativity, 
display a larger dispersion of the folding rates and a stronger 
topology-dependent kinetics 
than traditional, non cooperative G\={o} polymers. In a very recent study, 
Kaya and Chan suggested that the way thermodynamic cooperativity
is achieved may be as important as thermodynamic 
cooperativity {\it per se} in topology-dependent kinetics {\cite{KAYA1}}. 
By studying a modified G\={o} model, with many-body interactions,
the authors found folding rates, well correlated ($r=0.914$) with
$CO$, spanning a range two orders of magnitude larger than that of 
G\={o} models with additive contact energies. \par
The results for the modified G\={o} models and our current results for the 
MJ model shed light on our previous finding {\cite{PFN2}} of a 
particularly strong correlation ($r \approx 0.80$) between higher-$CO$ structures 
and longer logarithmic folding rates; These structures have a larger 
number of LR contacts that enhance the cooperativity of the 
folding transition. This cooperativity appears to be the essential ingredient of 
topology-dependent kinetics.\par

\clearpage

\begin{table*}
\caption{Summary of target properties with targets organized with increasing contact order parameter, $CO$. Targets $T_{0}$, $T_{1}$ and
$T_{2}$ constitute the low-$CO$ target set while targets $T_{3}$, $T_{4}$ and $T_{5}$ make the high-$CO$ target set. $<E>$ is the averaged trained energy and $\sigma$ its standard deviation , $\ln_{e}t$ is the logarithmic folding time, $Q$ is the fraction of native contacts with mean frequency $<\omega>$ (we specify the number of long-range (LR) native contacts) and $N_{nnat}$ is the number of non-native contacts with a marginal lifetime.}
\begin{ruledtabular}
\begin{tabular}{ccccccc} 
Target & $CO$ & $<E> \pm \sigma$ & $\ln t$ & $Q_{<\omega> \geq 0.5}$ &  $Q_{0.4 \leq <\omega> < 0.5}$ & $N_{nnat}(<\omega> \geq 0.1)$ \\ \hline
$T_{0}$ & $0.126$ & $ -25.80 \pm 0.03$ & $16.44 \pm 0.11$ & $0.25 (0 LR)$  & $0.19 (0 LR)$ & $22$ \\
$T_{1}$ & $0.127$ & $ -26.27 \pm 0.03$ & $14.99 \pm 0.13$ & $0.28 (1 LR)$ & $0.13 (1 LR)$ & $16$  \\
$T_{2}$ & $0.135$ & $ -25.78 \pm 0.04$ & $16.09 \pm 0.12$ & $0.19 (3 LR)$ & $0.13 (1 LR)$ & $25$  \\ \hline
$T_{3}$ & $0.241$ & $ -25.77 \pm 0.03$ & $16.83 \pm 0.16$ & $0.05 (0 LR)$        & $0.19 (5 LR)$ & $40$  \\
$T_{4}$ & $0.254$ & $ -25.11 \pm 0.03$ & $17.35 \pm 0.12$ & $0.11 (1 LR)$ & $0.16 (4 LR)$ & $38$  \\
$T_{5}$ & $0.259$ & $ -26.16 \pm 0.02$ & $17.59 \pm 0.12$ & $0.11 (2 LR)$ & $0.12 (4 LR)$ & $52$ 
\end{tabular}
\end{ruledtabular}
{\label{tab:tabno1}}
\end{table*}

\clearpage

\begin{table*}
\begin{ruledtabular}
\caption{Fraction of native contacts, $Q$, at consecutive intervals of backbone distance for targets $T_{1}$ and  $T_{5}$.}
\begin{tabular}{lcccccccc} 
Target & \multicolumn{8}{c}{backbone distance} \\ \cline{2-9}
& \multicolumn{1}{c}{$\lbrack 3, 8\lbrack$}
& \multicolumn{1}{c}{$\lbrack 8, 13\lbrack$}
& \multicolumn{1}{c}{$\lbrack 13, 18\lbrack$}
& \multicolumn{1}{c}{$\lbrack 18, 23\lbrack$}
& \multicolumn{1}{c}{$\lbrack 23, 28\lbrack$}
& \multicolumn{1}{c}{$\lbrack 28, 33\lbrack$}
& \multicolumn{1}{c}{$\lbrack 33, 38\lbrack$}
& \multicolumn{1}{c}{$\lbrack 38, 43\lbrack$}
\\\hline
$T_{1}$ & 0.49 & 0.18 & 0.19 & 0.07 & 0.05 & - & 0.02  & -   \\
$T_{5}$ & 0.23 & 0.04 & 0.09 & 0.19 & 0.14 & 0.05 & 0.14 & 0.13 
\end{tabular}
\label{tab:tabno2}
\end{ruledtabular}
\end{table*}

\clearpage

\begin{figure}
\centerline{\rotatebox{270}{\resizebox{8.6cm}{8.6cm}{\includegraphics{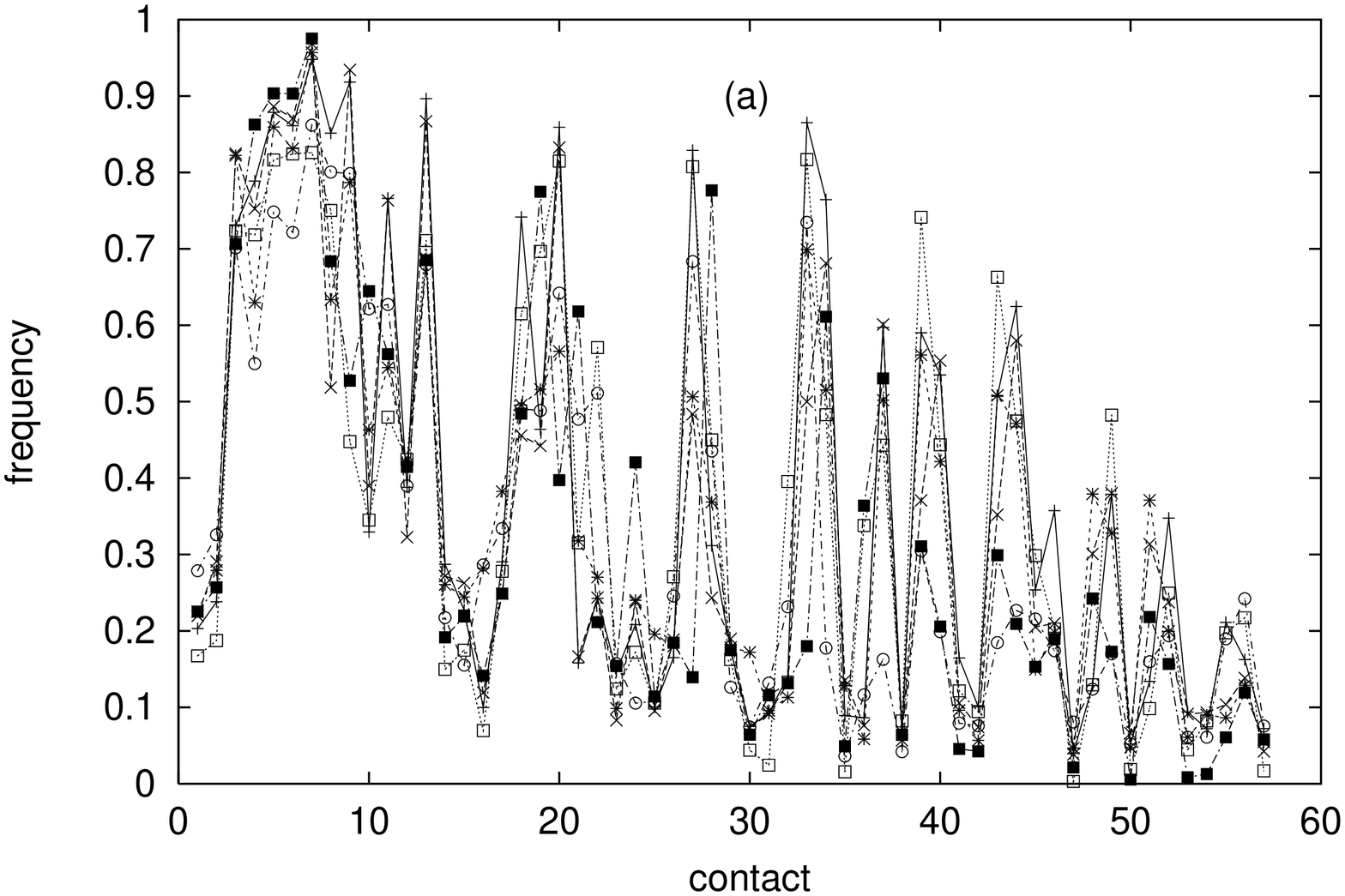}}}}
\end{figure}

\clearpage

\begin{figure}
\rotatebox{270}{\includegraphics{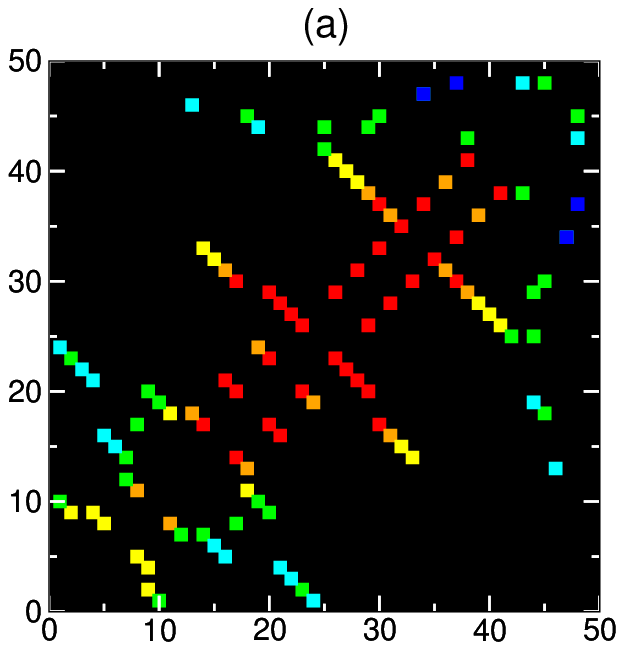}}
\rotatebox{270}{\includegraphics{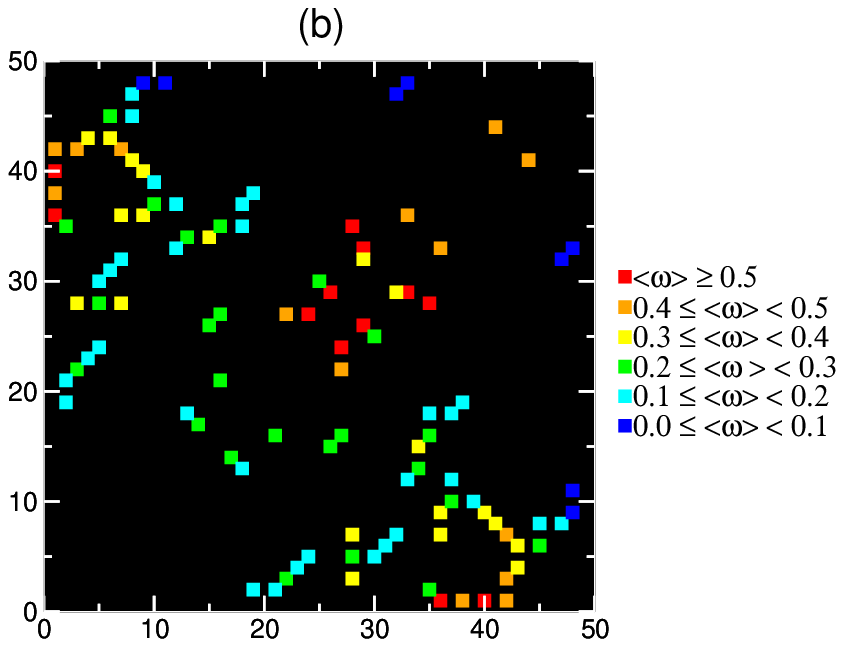}} 
\end{figure}

\begin{figure}
\centerline{\rotatebox{270}{\resizebox{8.6cm}{8.6cm}{\includegraphics{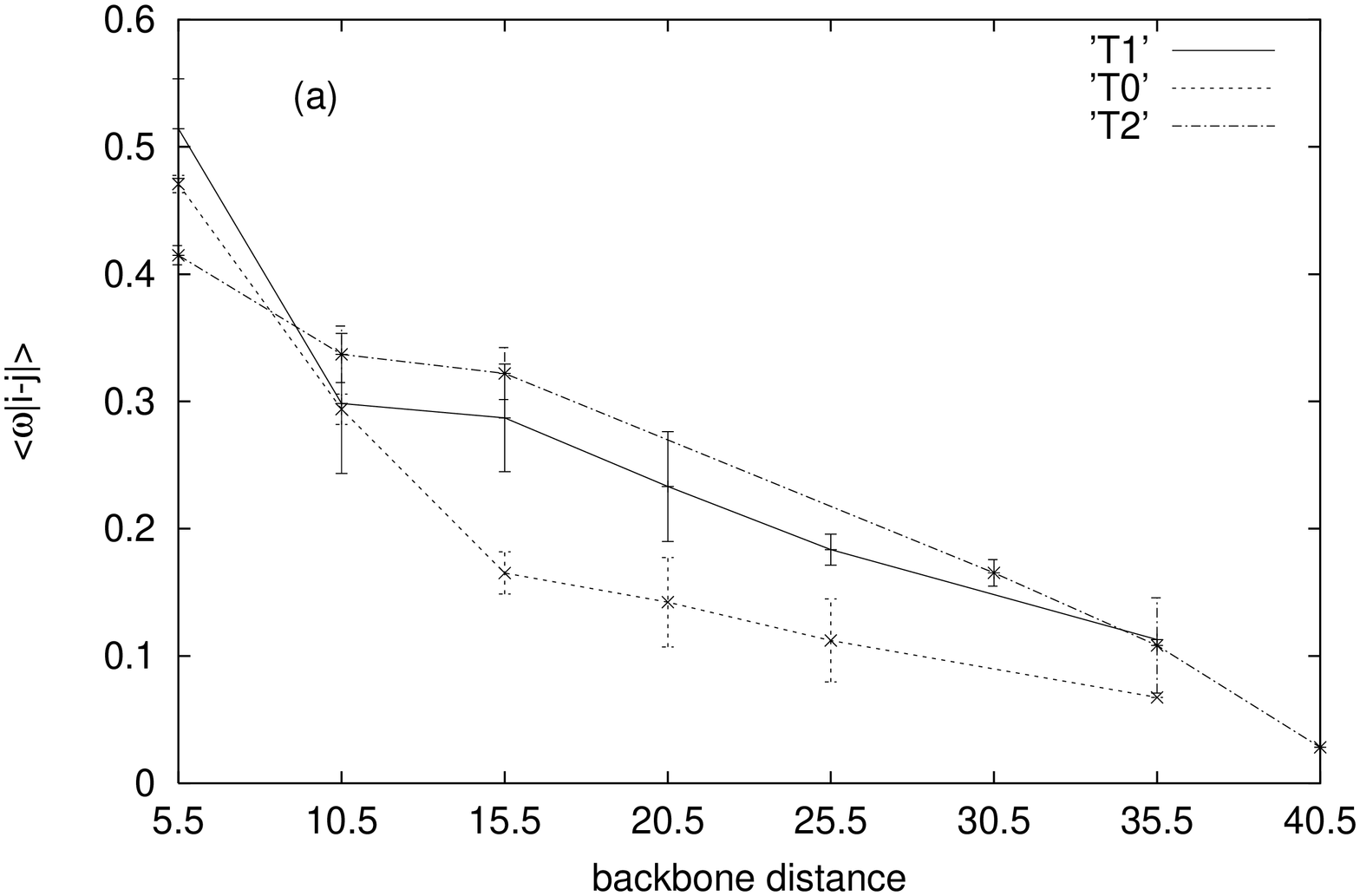}}}}
\end{figure}
\clearpage

\clearpage

\clearpage

\clearpage

\clearpage

\clearpage

\begin{acknowledgments}
P.F.N.F would like to thank Dr. A. Nunes for useful discussions and
Funda\c c\~ao para a Ci\^encia e Tecnologia for financial support through grant BPD10083/2002.
\end{acknowledgments}

\bibliography{resub_faisca}

\begin{thebibliography}{24}
\expandafter\ifx\csname natexlab\endcsname\relax\def\natexlab#1{#1}\fi
\expandafter\ifx\csname bibnamefont\endcsname\relax
  \def\bibnamefont#1{#1}\fi
\expandafter\ifx\csname bibfnamefont\endcsname\relax
  \def\bibfnamefont#1{#1}\fi
\expandafter\ifx\csname citenamefont\endcsname\relax
  \def\citenamefont#1{#1}\fi
\expandafter\ifx\csname url\endcsname\relax
  \def\url#1{\texttt{#1}}\fi
\expandafter\ifx\csname urlprefix\endcsname\relax\def\urlprefix{URL }\fi
\providecommand{\bibinfo}[2]{#2}
\providecommand{\eprint}[2][]{\url{#2}}

\bibitem[{\citenamefont{Jackson}(1997)}]{JACKSON}
\bibinfo{author}{\bibfnamefont{S.~E.} \bibnamefont{Jackson}},
  \bibinfo{journal}{Fold. Des.} \textbf{\bibinfo{volume}{3}},
  \bibinfo{pages}{81} (\bibinfo{year}{1997}).

\bibitem[{\citenamefont{Guijarro et~al.}(1998)\citenamefont{Guijarro, Morton,
  Plaxco, Campbell, and Dobson}}]{PLAXCO0}
\bibinfo{author}{\bibfnamefont{J.~I.} \bibnamefont{Guijarro}},
  \bibinfo{author}{\bibfnamefont{C.~J.} \bibnamefont{Morton}},
  \bibinfo{author}{\bibfnamefont{K.~W.} \bibnamefont{Plaxco}},
  \bibinfo{author}{\bibfnamefont{I.~D.} \bibnamefont{Campbell}},
  \bibnamefont{and} \bibinfo{author}{\bibfnamefont{C.~M.}
  \bibnamefont{Dobson}}, \bibinfo{journal}{J. Mol. Biol.}
  \textbf{\bibinfo{volume}{276}}, \bibinfo{pages}{657} (\bibinfo{year}{1998}).

\bibitem[{\citenamefont{Fersht}(2000)}]{FER}
\bibinfo{author}{\bibfnamefont{A.~R.} \bibnamefont{Fersht}},
  \bibinfo{journal}{Proc. Natl. Acad. Sci. USA} \textbf{\bibinfo{volume}{97}},
  \bibinfo{pages}{1525} (\bibinfo{year}{2000}).

\bibitem[{\citenamefont{Faisca and Ball}(2002{\natexlab{a}})}]{PFN}
\bibinfo{author}{\bibfnamefont{P.~F.~N.} \bibnamefont{Faisca}}
  \bibnamefont{and} \bibinfo{author}{\bibfnamefont{R.~C.} \bibnamefont{Ball}},
  \bibinfo{journal}{J. Chem. Phys.} \textbf{\bibinfo{volume}{116}},
  \bibinfo{pages}{7231} (\bibinfo{year}{2002}{\natexlab{a}}).

\bibitem[{\citenamefont{Kaya and Chan}(2003)}]{KAYA1}
\bibinfo{author}{\bibfnamefont{H.}~\bibnamefont{Kaya}} \bibnamefont{and}
  \bibinfo{author}{\bibfnamefont{H.~S.} \bibnamefont{Chan}},
  \bibinfo{journal}{Proteins} \textbf{\bibinfo{volume}{52}},
  \bibinfo{pages}{524} (\bibinfo{year}{2003}).

\bibitem[{\citenamefont{Wittung-Stafshede
  et~al.}(1999)\citenamefont{Wittung-Stafshede, Lee, Winkler, and
  Gray}}]{PERNILLA}
\bibinfo{author}{\bibfnamefont{P.}~\bibnamefont{Wittung-Stafshede}},
  \bibinfo{author}{\bibfnamefont{J.~C.} \bibnamefont{Lee}},
  \bibinfo{author}{\bibfnamefont{J.~R.} \bibnamefont{Winkler}},
  \bibnamefont{and} \bibinfo{author}{\bibfnamefont{H.~B.} \bibnamefont{Gray}},
  \bibinfo{journal}{Proc. Natl. Acad. Sci. USA} \textbf{\bibinfo{volume}{96}},
  \bibinfo{pages}{6857} (\bibinfo{year}{1999}).

\bibitem[{\citenamefont{van Nuland et~al.}(1998)\citenamefont{van Nuland,
  Chiti, Taddei, Raugei, Ramponi, and Dobson}}]{NIKOLAY}
\bibinfo{author}{\bibfnamefont{N.~A.~J.} \bibnamefont{van Nuland}},
  \bibinfo{author}{\bibfnamefont{F.}~\bibnamefont{Chiti}},
  \bibinfo{author}{\bibfnamefont{N.}~\bibnamefont{Taddei}},
  \bibinfo{author}{\bibfnamefont{G.}~\bibnamefont{Raugei}},
  \bibinfo{author}{\bibfnamefont{G.}~\bibnamefont{Ramponi}}, \bibnamefont{and}
  \bibinfo{author}{\bibfnamefont{C.~M.} \bibnamefont{Dobson}},
  \bibinfo{journal}{J. Mol. Biol.} \textbf{\bibinfo{volume}{283}},
  \bibinfo{pages}{883} (\bibinfo{year}{1998}).

\bibitem[{\citenamefont{Plaxco et~al.}(1998)\citenamefont{Plaxco, Simmons, and
  Baker}}]{PLAXCO}
\bibinfo{author}{\bibfnamefont{K.~W.} \bibnamefont{Plaxco}},
  \bibinfo{author}{\bibfnamefont{K.~T.} \bibnamefont{Simmons}},
  \bibnamefont{and} \bibinfo{author}{\bibfnamefont{D.}~\bibnamefont{Baker}},
  \bibinfo{journal}{J. Mol. Biol.} \textbf{\bibinfo{volume}{277}},
  \bibinfo{pages}{985} (\bibinfo{year}{1998}).

\bibitem[{\citenamefont{Du et~al.}(1999)\citenamefont{Du, Pande, Grosberg,
  Tanaka, and Shakhanovich}}]{DU}
\bibinfo{author}{\bibfnamefont{R.}~\bibnamefont{Du}},
  \bibinfo{author}{\bibfnamefont{V.~S.} \bibnamefont{Pande}},
  \bibinfo{author}{\bibfnamefont{A.~Y.} \bibnamefont{Grosberg}},
  \bibinfo{author}{\bibfnamefont{T.}~\bibnamefont{Tanaka}}, \bibnamefont{and}
  \bibinfo{author}{\bibfnamefont{E.}~\bibnamefont{Shakhanovich}},
  \bibinfo{journal}{J. Chem. Phys.} \textbf{\bibinfo{volume}{111}},
  \bibinfo{pages}{10375} (\bibinfo{year}{1999}).

\bibitem[{\citenamefont{Plaxco et~al.}(2000)\citenamefont{Plaxco, Simmons,
  Ruczinski, and Baker}}]{PLAXCO2}
\bibinfo{author}{\bibfnamefont{K.~W.} \bibnamefont{Plaxco}},
  \bibinfo{author}{\bibfnamefont{K.~T.} \bibnamefont{Simmons}},
  \bibinfo{author}{\bibfnamefont{I.}~\bibnamefont{Ruczinski}},
  \bibnamefont{and} \bibinfo{author}{\bibfnamefont{D.}~\bibnamefont{Baker}},
  \bibinfo{journal}{Biochemistry} \textbf{\bibinfo{volume}{39}},
  \bibinfo{pages}{11177} (\bibinfo{year}{2000}).

\bibitem[{\citenamefont{Faisca and Ball}(2002{\natexlab{b}})}]{PFN2}
\bibinfo{author}{\bibfnamefont{P.~F.~N.} \bibnamefont{Faisca}}
  \bibnamefont{and} \bibinfo{author}{\bibfnamefont{R.~C.} \bibnamefont{Ball}},
  \bibinfo{journal}{J. Chem. Phys.} \textbf{\bibinfo{volume}{117}},
  \bibinfo{pages}{8587} (\bibinfo{year}{2002}{\natexlab{b}}).

\bibitem[{\citenamefont{Jewett et~al.}(2003)\citenamefont{Jewett, Pande, and
  Plaxco}}]{JEWETT}
\bibinfo{author}{\bibfnamefont{A.~I.} \bibnamefont{Jewett}},
  \bibinfo{author}{\bibfnamefont{V.~S.} \bibnamefont{Pande}}, \bibnamefont{and}
  \bibinfo{author}{\bibfnamefont{K.~W.} \bibnamefont{Plaxco}},
  \bibinfo{journal}{J. Mol. Biol.} \textbf{\bibinfo{volume}{326}},
  \bibinfo{pages}{247} (\bibinfo{year}{2003}).

\bibitem[{\citenamefont{Mirny and Shakhnovich}(2001)}]{SFERMI}
\bibinfo{author}{\bibfnamefont{L.}~\bibnamefont{Mirny}} \bibnamefont{and}
  \bibinfo{author}{\bibfnamefont{E.~I.} \bibnamefont{Shakhnovich}},
  \bibinfo{journal}{Proc. of the International School of Physics Enrico Fermi:
  Protein Folding Evolution and Design, IOS Press} p.~\bibinfo{pages}{37}
  (\bibinfo{year}{2001}).

\bibitem[{\citenamefont{Shakhnovich and Gutin}(1993)}]{SG}
\bibinfo{author}{\bibfnamefont{E.~I.} \bibnamefont{Shakhnovich}}
  \bibnamefont{and} \bibinfo{author}{\bibfnamefont{A.~M.} \bibnamefont{Gutin}},
  \bibinfo{journal}{Proc. Natl. Acad. Sci. USA} \textbf{\bibinfo{volume}{90}},
  \bibinfo{pages}{7195} (\bibinfo{year}{1993}).

\bibitem[{\citenamefont{Shakhnovich}(1994)}]{SG2}
\bibinfo{author}{\bibfnamefont{E.~I.} \bibnamefont{Shakhnovich}},
  \bibinfo{journal}{Phys. Rev. Lett.} \textbf{\bibinfo{volume}{72}},
  \bibinfo{pages}{3907} (\bibinfo{year}{1994}).

\bibitem[{\citenamefont{Sali et~al.}(1994)\citenamefont{Sali, Shakhnovich, and
  Karplus}}]{SALI}
\bibinfo{author}{\bibfnamefont{A.}~\bibnamefont{Sali}},
  \bibinfo{author}{\bibfnamefont{E.~I.} \bibnamefont{Shakhnovich}},
  \bibnamefont{and} \bibinfo{author}{\bibfnamefont{M.}~\bibnamefont{Karplus}},
  \bibinfo{journal}{Nature (London)} \textbf{\bibinfo{volume}{369}},
  \bibinfo{pages}{248} (\bibinfo{year}{1994}).

\bibitem[{\citenamefont{Tiana and Broglia}(2001)}]{TIANA}
\bibinfo{author}{\bibfnamefont{G.}~\bibnamefont{Tiana}} \bibnamefont{and}
  \bibinfo{author}{\bibfnamefont{R.~A.} \bibnamefont{Broglia}},
  \bibinfo{journal}{J. Chem. Phys.} \textbf{\bibinfo{volume}{114}},
  \bibinfo{pages}{2503} (\bibinfo{year}{2001}).

\bibitem[{\citenamefont{Abkevich et~al.}(1995)\citenamefont{Abkevich, Gutin,
  and Shakhnovich}}]{JMBS}
\bibinfo{author}{\bibfnamefont{V.~I.} \bibnamefont{Abkevich}},
  \bibinfo{author}{\bibfnamefont{A.~M.} \bibnamefont{Gutin}}, \bibnamefont{and}
  \bibinfo{author}{\bibfnamefont{E.~I.} \bibnamefont{Shakhnovich}},
  \bibinfo{journal}{J. Mol. Biol.} \textbf{\bibinfo{volume}{252}},
  \bibinfo{pages}{460} (\bibinfo{year}{1995}).

\bibitem[{\citenamefont{Miyazawa and Jernigan}(1985)}]{MJ}
\bibinfo{author}{\bibfnamefont{S.}~\bibnamefont{Miyazawa}} \bibnamefont{and}
  \bibinfo{author}{\bibfnamefont{R.}~\bibnamefont{Jernigan}},
  \bibinfo{journal}{Macromolecules} \textbf{\bibinfo{volume}{18}},
  \bibinfo{pages}{534} (\bibinfo{year}{1985}).

\bibitem[{\citenamefont{Metropolis et~al.}(1953)\citenamefont{Metropolis,
  Rosenbluth, Rosenbluth, Teller, and Teller}}]{METROPOLIS}
\bibinfo{author}{\bibfnamefont{N.}~\bibnamefont{Metropolis}},
  \bibinfo{author}{\bibfnamefont{A.~W.} \bibnamefont{Rosenbluth}},
  \bibinfo{author}{\bibfnamefont{M.~N.} \bibnamefont{Rosenbluth}},
  \bibinfo{author}{\bibfnamefont{A.~H.} \bibnamefont{Teller}},
  \bibnamefont{and} \bibinfo{author}{\bibfnamefont{E.}~\bibnamefont{Teller}},
  \bibinfo{journal}{J. Chim. Phys.} \textbf{\bibinfo{volume}{21}},
  \bibinfo{pages}{1087} (\bibinfo{year}{1953}).

\bibitem[{\citenamefont{Landau and Binder}(2000)}]{BINDER}
\bibinfo{author}{\bibfnamefont{D.~P.} \bibnamefont{Landau}} \bibnamefont{and}
  \bibinfo{author}{\bibfnamefont{K.}~\bibnamefont{Binder}},
  \bibinfo{journal}{Cambridge Univesrity Press} p. \bibinfo{pages}{1087}
  (\bibinfo{year}{2000}).

\bibitem[{\citenamefont{Saitoh et~al.}(1993)\citenamefont{Saitoh, Nakai, and
  Nishikawa}}]{CMAP}
\bibinfo{author}{\bibfnamefont{S.}~\bibnamefont{Saitoh}},
  \bibinfo{author}{\bibfnamefont{T.}~\bibnamefont{Nakai}}, \bibnamefont{and}
  \bibinfo{author}{\bibfnamefont{K.}~\bibnamefont{Nishikawa}},
  \bibinfo{journal}{Proteins: Struct. Funct. and Genet.}
  \textbf{\bibinfo{volume}{15}}, \bibinfo{pages}{191} (\bibinfo{year}{1993}).

\bibitem[{\citenamefont{Lazaridis and Karplus}(1997)}]{KPLUS}
\bibinfo{author}{\bibfnamefont{T.}~\bibnamefont{Lazaridis}} \bibnamefont{and}
  \bibinfo{author}{\bibfnamefont{M.}~\bibnamefont{Karplus}},
  \bibinfo{journal}{Science} \textbf{\bibinfo{volume}{278}},
  \bibinfo{pages}{1928} (\bibinfo{year}{1997}).

\bibitem[{\citenamefont{Plaxco and Makarov}(2003)}]{PLAXCO3}
\bibinfo{author}{\bibfnamefont{K.~W.} \bibnamefont{Plaxco}} \bibnamefont{and}
  \bibinfo{author}{\bibfnamefont{D.~E.} \bibnamefont{Makarov}},
  \bibinfo{journal}{Protein Sci.} \textbf{\bibinfo{volume}{12}},
  \bibinfo{pages}{17} (\bibinfo{year}{2003}).

\end{thebibliography}

\end{document}